# Visitor Pattern: Implementation of Enquiry Pattern for Internet Banking

J. Prabavadhi

PG Student, Department of CSE, Sri Manakula Vinayagar Engineering College, Puducherry

A.Meiappane

Research Scholar, , Pondicherry University, Puducherry

Dr. V. Prasanna Venkatesan

Associate Professor, Pondicherry University, Puducherry

auromei@yahoo.com          prasanna_v@yahoo.com

## Abstract

*This paper brings out the design patterns according to the various services involved in internet banking. The Pattern oriented Software Architecture uses the pattern which eliminates the difficulty of reusability in a particular context. The patterns are to be designed using BPM (Business Process Model) for effective cross cutting on process level. For implementing the above said BPM, the Internet banking has been taken to implement the pattern into it. The Analysis and identification of various processes in Internet Banking have been done, to identify the effective cross cutting features. With this process the pattern has been designed, as a reusability component to be used by the Software Architect. The pattern help us to resolve recurring problems constructively and based on proven solutions and also support us in understanding the architecture of a given software system. Once the model is finalized by analyzing we found enquiry pattern as the visitor pattern and implement the pattern.*

## 1. Introduction

In software engineering, a design pattern is a general reusable solution to a commonly occurring problem within a given context in software design. A design pattern is not a finished design that can be transformed directly into code. It is a description or template for how to solve a problem that can be used in many different situations and it is also called as a blue-print of how to solve a problem.    It is used to determine implementation faster and make code more readable to other programmers the design pattern is divided into three types: creational, structural, and behavioral. *Creational patterns* create objects for you rather than having you instantiate objects directly. This gives your program more flexibility in deciding which objects need to be created for a given case. *Structural patterns* help you compose groups of objects into larger structures, such as complex user interfaces or accounting data. *Behavioral patterns* help you define the communication between objects in your system and how the flow is controlled in a complex program [14].

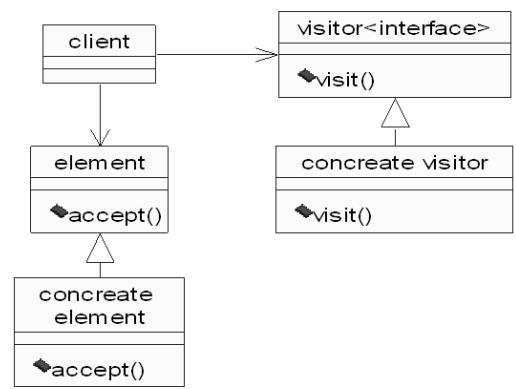

Figure 1.Structure of Visitor Pattern

The Visitor pattern has been concentrated here which is a behavioral pattern. The visitor design pattern is a way of separating an algorithm from an object structure on which it operates. A practical result of this separation is the ability to add new operations to existing object structures without modifying those structures.

## 1.1. Software Architecture

Software architecture is the fundamental organization of a system, embodied in its components, their relationships to each other and to the environment, and the principles guiding its design and evolution.

## 1.2. Patterns

A particular recurring design problem that arises in specific design contexts, and presents a well-proven generic scheme for its solution. The solution scheme is specified by describing its constituent components, their responsibilities and relationships, and the ways in which they collaborate. Each pattern is a three-part rule, which expresses a relation between a certain context, a problem and a solution. Patterns can be applied to many different areas of human endeavor, including software development [14].

## 1.3. Architectural pattern

It is Standard design and it Express a fundamental structural organization for software system and provide set of predefined sub system and includes rules and guidelines for organizing the relationships between them.

## 1.4. Usage of patterns

- Comprehend existing systems; customize systems to fit user needs; and construct new systems.
- Identify and specify higher level abstractions
- Reinforce an architectural view of the system
- Explicitly address non-functional properties
- High availability and minimization of business risk
- They help in the construction of complex and heterogeneous software architectures
- They help to manage complexity

## 2. Pattern Mining

Pattern mining is same as the data mining. Mining the pattern in internet banking by using the BPM (See Figure 2). It is used to representing process involved in internet banking and is used to analyze the process then apply the crosscutting concerns approach in this model. The crosscutting concerns are the Aspect of program which affects the other concern. These concerns often cannot be cleanly decomposed from the rest of the system in both design and implementation and can the result in either scattering, tangling. The crosscutting concern is used to improve the modularity of the system.

In internet bank has lots of services and we analysis this services and find the crosscutting in the process level. The services are the account, Third party transfer, bill payment, credit card, debit card, Mutual fund.

Use of the crosscutting concern we found the enquiry pattern. The different types of enquiry available in the internet banking, we analysis the services and find the crosscutting in enquiry .The different types of enquiry are view account statement, balance, mini statement and in credit card it have the same view options are the statement, balance, mini statement. We put these services under the enquiry pattern and this pattern is like the visitor design pattern. The pattern is used to increase the reusability and reduce the number of code and easily add new services to the internet bank with the small modification.

## 3. Pattern construction

A pattern for software architecture describes a particular recurring design problem that arises in specific design contexts and presents a well-proven generic scheme for its solution. The solution scheme is specified by describing its constituent components, their responsibilities and relationships, and the ways in which they collaborate."

Each pattern is a three-part rule, which expresses a relation between a certain context, a certain system of forces which occurs repeatedly in that context, and a certain software configuration which allows these forces to resolve themselves. Context section describes the situation in which the design problem arises. Problem section describes the problem that arises repeatedly in the context. And finally, solution section describes a proven solution to the problem. Mining the enquiry pattern for internet banking. This pattern is like the visitor pattern. The pattern is constructed by using the elements are Name, Intent, Problem, and Solution (see Figure 3).

## 3.1. Implementation of Enquiry Pattern

**Intent:** This pattern is used to view different account type by different services through online by separating the account type from the services.

**Problem:** how can easily add new services to the account type without modify the existing code.

**Solution:** The Enquiry is the type of the visitor pattern. Two interfaces involved: Account and Services. The Account system is completely independent."How to get service" tries to add new operation to Account system. It is done by adding another interface Services and parameter zing Account interface in the abstract method visit (composite pattern). The "how to get" classes implement services interface and make a connection with Account system.

In this implementation two interfaces are used one is the account type (visitor) and services (element). The concrete visitors are the account enquiry, credit card enquiry. Concrete elements are the statement visitor, balance visitor, mini statement visitor. The implementation mentioned below for the enquiry visitor pattern.

```
public interface Accounts {
public void visit(Services s);
}

public class Statement implements Services {
public int service()
  {
  System.out.println("Statement called");
  return 1;
  }
}
public class AccountEnquiry implements Accounts{
    public void visit(Services s)
  {
  System.out.println("AccountEnquiry called");
  i=s.service();
  System.out.println(i);
  if(i==1)
  {
    new accstatementenq().setVisible(true);
  }
else if(i==2)
  {
  new accbalanceenq().setVisible(true);;
}
else
  {
    new accministatement().setVisible(true);;
}
}
```

```
}
}

public class ServiceSelect {
public Accounts getType(int i){
        System.out.println("Test 2");
   switch (i){
      case 0: return new AccountEnquiry();
      case 1: return new CreditCardEnquiry();
      default: return null;
           }
}
public Services getService(int j){
  System.out.println("Test 3");
    switch (j){
      case 0: return new Statement();
      case 1: return new Balance();
      case 2: return new MiniStatement();
      default: return null;
          }
}
}
```

## 4. Conclusions

The visitor pattern for internet banking has been presented over here. These patterns focus mainly on ways to solve usual problem. They provide hints to the Web application designer in order to make these applications more usable and effective both from the point of customers and designer of web application. These patterns help to improve the design and reusability. Further the research is preceded on mining some more patterns for internet banking which will be a reusable component for internet banking at a particular context.

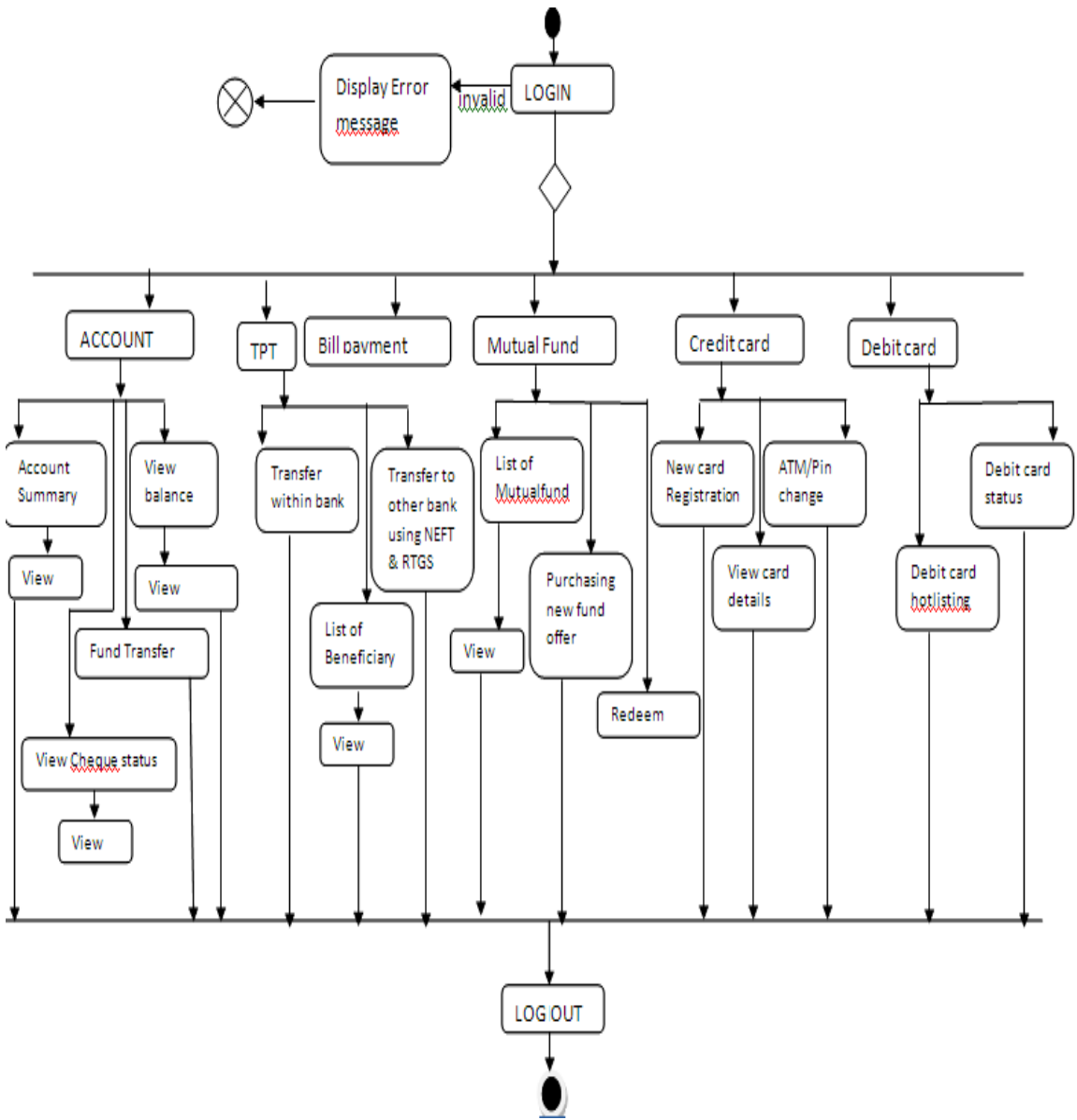

Figure 2. Business Process Model for Internet Banking

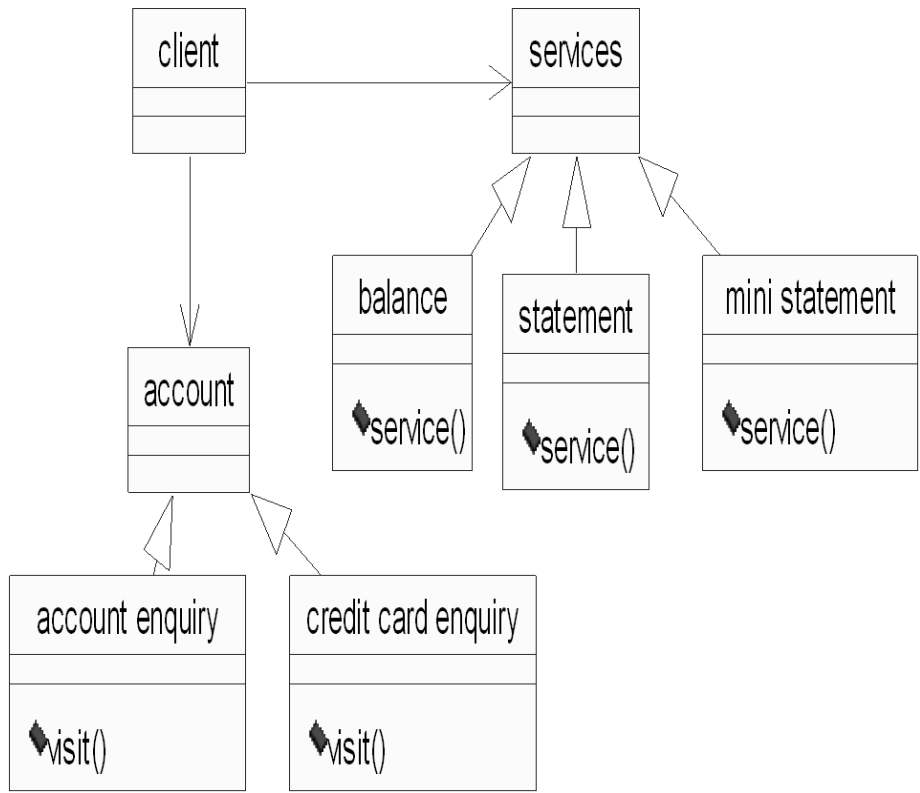

Figure 3. Visitor Pattern for Enquiry

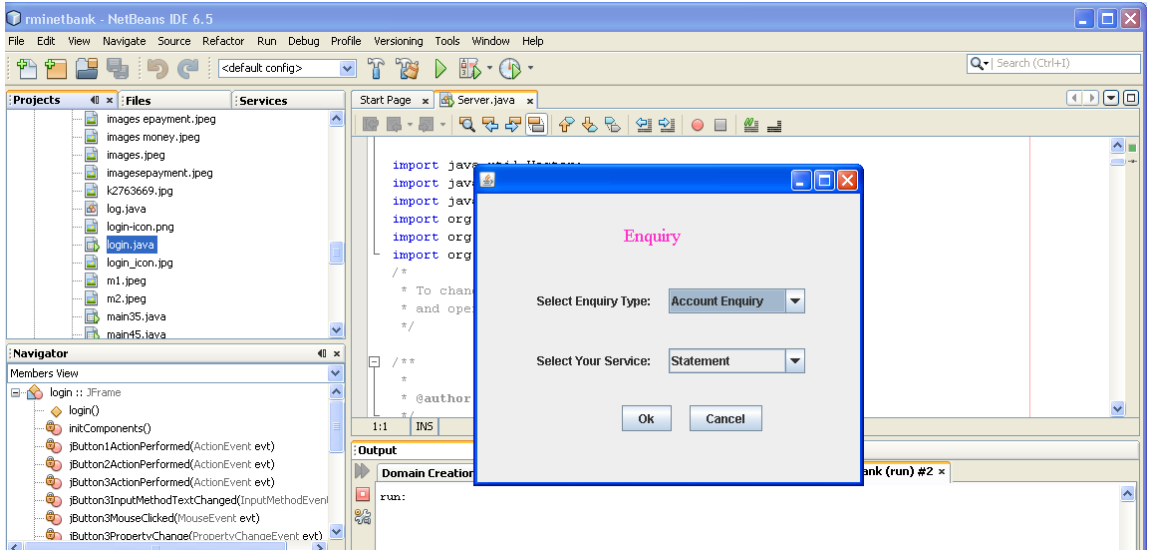

Figure 4. Enquiry

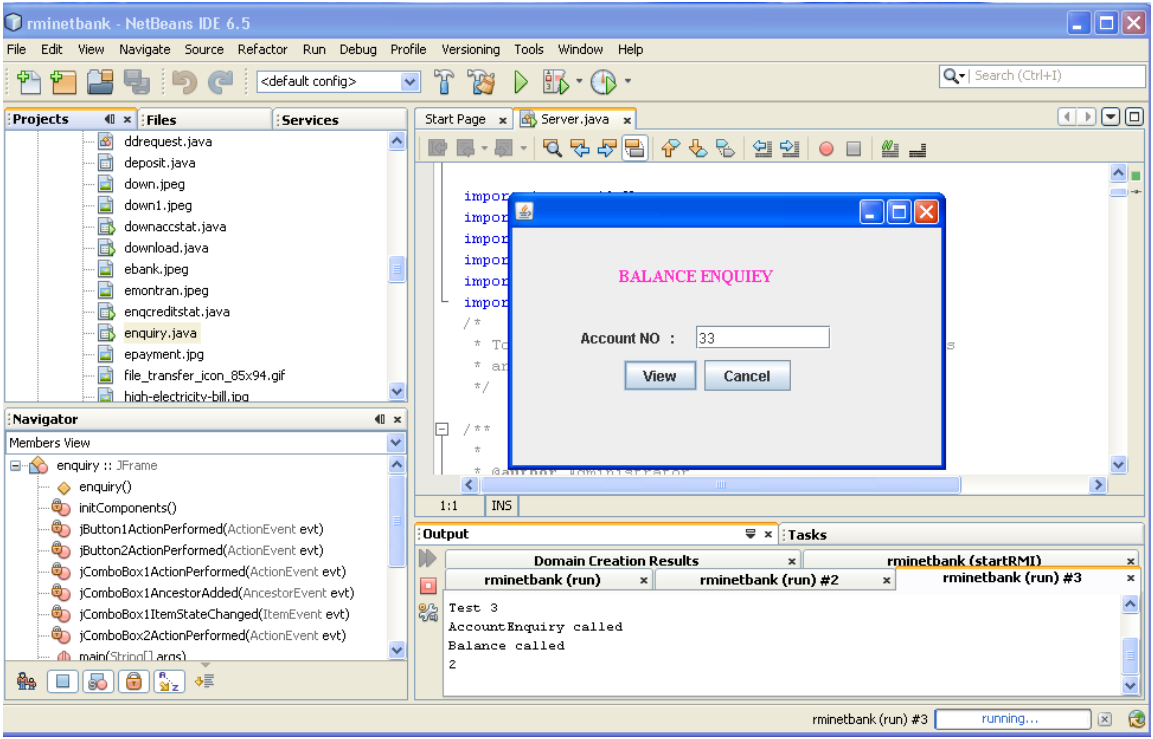

Figure 5. Balance Enquiry

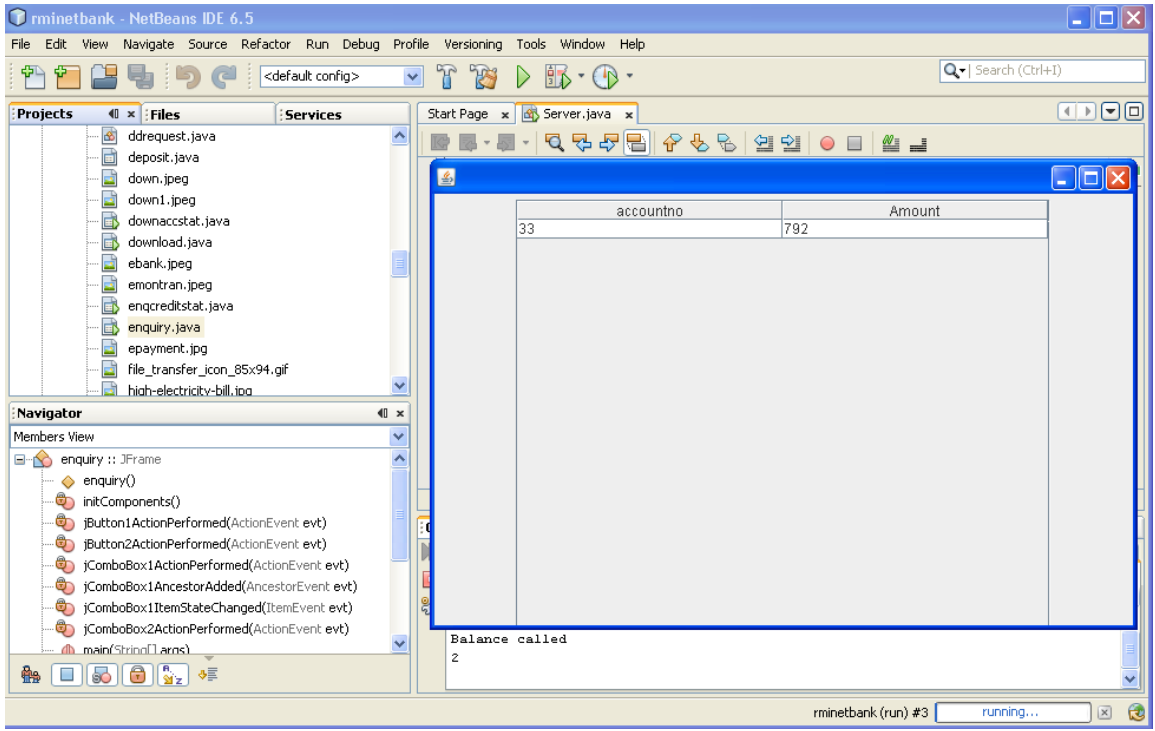

Figure 6. View Balance Detail